\documentclass{iaus}
\usepackage{graphicx}

\title[Stellar activity, differential rotation, and exoplanets] 
{Stellar activity, differential rotation, \\
 and exoplanets}

\author[A.~F.~Lanza]   
{A.~F.~Lanza}

\affiliation{INAF-Osservatorio Astrofisico di Catania, Via S.~Sofia, 78, 95123 Catania, Italy \\ email: {\tt nuccio.lanza@oact.inaf.it} }

\pubyear{2010}
\volume{273}  
\pagerange{--}
\jname{Physics of Sun and Star Spots}
\editors{D.~P.Choudhari \& K.~G.~Strassmeier, eds.}

\begin{document}

\maketitle

\begin{abstract}
The photospheric spot activity of some of the stars with transiting planets discovered by the CoRoT space experiment is reviewed.  Their out-of-transit light modulations are fitted by  a spot model previously tested with the total solar irradiance variations. This approach allows us to study the longitude distribution of the spotted area and its variations versus time during the five months of a typical CoRoT time series. The migration of the spots in longitude provides a lower limit for the surface differential rotation, while the variation of the total spotted area can be used to search for short-term  cycles akin the solar Rieger cycles. The possible impact of a close-in giant planet on stellar activity is also discussed.  
\end{abstract}

\firstsection 
\section{Introduction}
Magnetic fields in the Sun and late-type stars produce active regions in the photosphere by affecting the transfer of energy and momentum in the outermost convective layers. These features consist of dark spots, bright faculae, and an enhanced network  of magnetic flux tubes with radii down to about 100 km and kG field strength.  Solar active regions can be studied in details thanks to a spatial resolution down to $\sim 100$~km and a time resolution better than $\sim 1$~s. 

In distant stars, we lack spatial resolution (with the exception of the  supergiant $\alpha$ Orionis, see \cite[Dupree 2010]{Dupree10}), so we must apply indirect techniques to map their photospheres. The most successful is Doppler Imaging that produces a two-dimensional map of the surface from a sequence of high-resolution line profiles, sampling the different rotation phases of the star (see, e.g., \cite[Strassmeier 2009, 2010]{Strassmeier09,Strassmeier10}). Its application requires that the rotational broadening of the spectral lines exceeds  the  macroturbulence by at least a factor of $4-5$,  implying a minimum $ v \sin i \sim 10-15$~km~s$^{-1}$  in dwarf stars. Therefore, late-type stars that rotate as slowly as the Sun cannot be imaged by this technique. For those stars, we derive  the distribution of brightness inhomogeneities from the rotational  modulation of the flux produced by cool spots and bright faculae that come into and out of view as the star rotates. By comparing successive rotations, it is possible to study the time evolution of the active regions. 

Another method  exploits the tiny light modulations produced by the occultation of starspots by a transiting giant planet moving in front of the disc of its star (e.g. \cite[Silva-Valio et al. 2010; Wolter et al.  2009]{SilvaValioetal10,Wolteretal09}). It is a specialized version of the eclipse mapping technique, developed for  the active components of close binary stars (e.g., \cite[Collier Cameron 1997; Lanza et al.  1998)]{CollierCameron97,Lanzaetal98}. Its application will be reviewed by \cite[Silva-Valio (2010)]{SilvaValio10}.
Stellar differential rotation and activity cycles can be studied also through  techniques of time series analysis applied to  sequences of  seasonal optical photometry; see, e.g., \cite[Messina \& Guinan (2003); Koll\'ath \& Ol\'ah (2009); and Ol\'ah (2010)]{MessinaGuinan03,KollathOlah09,Olah10}.

In this review, I shall briefly report on the  modelling of the light curves of some planet-hosting stars as observed by the space experiment CoRoT.

\section{Space-borne optical photometry}
Space-borne optical photometry has provided  time series  for several late-type stars spanning from several days up to several months thanks to the space missions MOST \cite[(Walker et al. 2003)]{Walkeretal03}, CoRoT \cite[(Auvergne et al. 2009)]{Auvergneetal09}, and Kepler \cite[(see, e.g., Haas et al. 2010; Ciardi et al. 2010)]{Haasetal10,Ciardietal10}. 

MOST (the Microvariability and Oscillation of Stars satellite) has a telescope of 15~cm aperture that can observe a given target for up to $40-60$ days reaching a photometric precision of $50-100$ ppm (parts per million) on the brightest stars ($V \leq 4$). It has observed $\epsilon$~Eridani \cite[(Croll et al. 2006)]{Crolletal06} and k$^1$~Ceti \cite[(Walker et al.  2007)]{Walkeretal07} whose light curves have been modelled to extract information on stellar differential rotation. 

CoRoT (Convection, Rotation and Transits) is a space experiment devoted to asteroseismology and the search for extrasolar planets through the method of transits. With its 27-cm aperture telescope, it can observe up to 12,000 targets per field for intervals of 150 days searching for transit signatures in the  light curves. Its white-band light curves have a sampling  of 32 or 512~s and a bandpass ranging from 300 to 1100~nm.  A photometric accuracy of $\sim 100$~ppm is achieved for a G or K-type star of $V \sim 12$ by integrating the flux over individual orbits of the satellite. CoRoT provides some chromatic information on the light variability of its brightest targets ($V < 14.5$). 

Kepler, launched in March 2009,  has a telescope of 95~cm aperture and is continuously monitoring $\sim 100,000$  dwarfs in a fixed field of view for a time interval of at least 3.5 years, searching for planetary transits. Its accuracy reaches $\sim 30$~ppm in 1 hour integration on a G2V target of $V \sim 12$. 

Stars with transiting giant planets allow us to derive  the inclination of the stellar rotation axis, which is important to modelling  their light curves for stellar activity studies (see, e.g., \cite[Mosser et al.  2009]{Mosseretal09}). Specifically, by fitting the transit light curve, we can measure the inclination of the planetary orbit along the line of sight, which is equal to the inclination of the stellar rotation axis if the spin  and the orbital angular momentum are aligned. This hypothesis can be tested with the so-called Rossiter-McLaughlin effect, i.e., the apparent anomaly in the radial velocity of the star observed during the planetary transit that allows us to measure the angle between the projections of the stellar spin and  the orbital angular momentum on the plane of the sky \cite[(Winn et al. 2005)]{Winnetal05}. 
Another advantage of  stars with transiting planets is that their fundamental parameters have  been well determined   because accurate stellar masses and radii are needed to derive accurate planetary parameters from the transit modelling.

\section{Light curve modelling}
Spot models of the light curves of  late-type  stars observed by MOST and CoRoT have already been published, while  the modelling of Kepler light curves has just started \cite[(e.g., Brown et al. 2010)]{Brownetal10}.  MOST time series were exploited to measure  stellar differential rotation  by fitting the wide-band light modulation with a few circular spots with  fixed contrast. Thanks to such assumptions, it was possible to derive  the latitudes of the spots and measure the variation of the angular velocity vs. latitude
\cite[(Croll et al. 2006; Walker et al. 2007)]{Crolletal06,Walkeretal07}.
 For a generalization of that approach with the CoRoT light curves, see \cite[Mosser et al. (2009)]{Mosseretal09}.   

In the  Sun, active regions consist not only of cool spots but also of bright faculae whose contrast is maximum close to the limb and minimum at the disc centre. Moreover, the optical variability of the Sun is dominated by   several active regions at the same time  making a model based on a few spots poorly suitable to reproduce its active region pattern.

\cite[Lanza et al. (2007)]{Lanzaetal07} used the time series of the Total Solar Irradiance (TSI, e.g., \cite[Fr\"ohlich \& Lean 2004]{FrohlichLean04}) as a proxy for the solar optical light curve to test different modelling approaches assuming that the active regions consist of dark spots and bright faculae with fixed contrasts and in a fixed area proportion. A model with a continuous distribution of active regions and the maximum entropy regularization, to warrant the uniqueness and stability of the solution, is the most suitable  and reproduces the distribution of the area of the sunspot groups vs.  longitude with a  resolution better that $\sim 50^{\circ}$, as well as the variation of the total spotted area vs. time. It allows a highly accurate reproduction of the TSI variations with a typical standard deviation of the residuals of $\sim 30-35$ ppm for time intervals of 14 days. However, the value of the faculae-to-spotted area ratio is a critical parameter  because it affects the derived distribution of the active regions vs. longitude  (see, \cite[Lanza et al. 2007]{Lanzaetal07}, for a detailed discussion). 

The maximum entropy spot model tested in the case of the Sun has been  applied to CoRoT light curves  to derive the distribution of the stellar active regions vs. longitude and the variation of their total area vs. time. In general, information on the latitudes of stellar active regions cannot be extracted from a one-dimensional data set such as an optical light curve. Moreover, since the inclination of the rotation axis  of the stars with transiting planets is generally close to $90^{\circ}$, it is impossible to constrain the spot latitudes because the duration of the transit of a spot across the stellar disc  is independent of its latitude. 

\section{Results from  CoRoT light curves}
\cite[Lanza et al. (2009a)]{Lanzaetal09a} model the out-of-transit light curve of CoRoT-2, a G7V star with  a giant  planet with a mass of 3.3 Jupiter masses and an orbital period of 1.743~days \cite[(Alonso et al. 2008; Bouchy et al.  2008)]{Alonsoetal08,Bouchyetal08}. Since the spot pattern is evolving rapidly, they model individual intervals of 3.15 days along a sequence of 142~days.  The star has a light curve amplitude of 0.06 mag, i.e., about 20 times that of the Sun at the maximum of the 11-yr cycle. Solar-like contrasts for the spots and the  faculae are adopted. 

\begin{figure}[]
\begin{center}
 \includegraphics[width=1.7in,height=3.8in,angle=90]{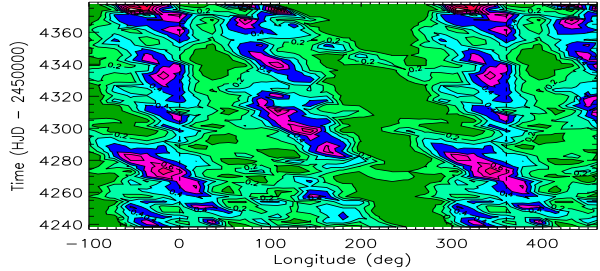} 
 \caption{The isocontours of the spot filling factor vs. longitude and time for CoRoT-2 after \cite[Lanza et al. (2009a)]{Lanzaetal09a}. The longitude reference frame rotates with the star with a period of 4.5221 days and the longitude increases in the same direction of the stellar rotation. The longitude scale has been extended beyond $0^{\circ}$ and $360^{\circ}$ (marked by the vertical dashed lines) to help follow the migration of the spots. In the electronic version,  different colours indicate different relative filling factors of the starspots; from the minimum to the maximum: green, light green, light blue, blue, light pink, pink, and red. }
   \label{lanza_fig1}
\end{center}
\end{figure}

The distribution of the spotted area vs. longitude and time is plotted in Fig.~4 of \cite[Lanza et al. (2009a)]{Lanzaetal09a}, here reproduced in Fig.~\ref{lanza_fig1}. The active regions are mainly found within two active longitudes, initially separated by $\sim 180^{\circ}$. The longitude initially around  $0^{\circ}$ does not migrate, i.e., it rotates with the same period as the adopted reference frame, while the other longitude, initially around $180^{\circ}$, migrates  backward, i.e., it rotates  slower than the reference frame by $\sim 0.9$ percent. Individual active regions  also migrate backward as they evolve, i.e., they rotate slower than the active longitude to which they belong,  with a maximum difference of $\approx 2$ percent. The relative migration of the active longitudes can be interpreted as a consequence of their different latitudes on a differentially rotating star,  yielding a lower limit of 0.9 percent for the amplitude of the differential rotation \cite[(Lanza et al. 2009a)]{Lanzaetal09a}. On the other hand, the backward migration of the individual active regions can be regarded as  analogous to the braking of the rotation of solar active regions as they evolve because the relative amplitude of the angular velocity variation is remarkably similar (e.g., \cite[Zappal\`a \& Zuccarello 1991; Sch\"ussler \& Rempel  2005]{ZappalaZuccarello91,SchuesslerRempel05}). 

Other authors have suggested a greater amplitude for the differential rotation of CoRoT-2, up to $\sim 8$ percent, from the migration of individual active regions \cite[(see Fr\"ohlich et al. 2009; Huber et al.  2010)]{Froehlichetal09,Huberetal10}. \cite[Savanov (2010)]{Savanov10} notices that  the active regions appear in an  alternate way in the two active longitudes, suggesting a short-term flip-flop phenomenon,  reminiscent of the flip-flop cycles in some active stars that, however, have timescales of several years \cite[(Berdyugina \& Tuominen 1998; Berdyugina 2005)]{BerdyuginaTuominen98,Berdyugina05}.  

The variation of the total spotted area shows remarkable oscillations with a cycle of $ \sim 29$ days (see Fig.~6 of \cite[Lanza et al. 2009a]{Lanzaetal09a}). In the Sun, short-term  oscillations of the total spotted area have been observed close to the maximum of some of the 11-yr cycles and are called Rieger cycles. They have periods around 160 days, i.e., about five times longer than the cycles observed in CoRoT-2 \cite[(Oliver et al. 1998; Krivova \& Solanki 2002; Zaqarashvili et al.  2010)]{Oliveretal98,KrivovaSolanki02,Zaqarashvilietal10}. \cite[Lou (2000)]{Lou00} suggests that they may be due to hydromagnetic Rossby-type waves trapped in the solar convection zone. Since the wave frequency is proportional to the rotation frequency of the star, the expected period is close to that observed in CoRoT-2 because this star rotates five times faster than the Sun.

The approach introduced for CoRoT-2 has been applied to  other late-type planet-hosting stars, viz. CoRoT-4 \cite[(Lanza et al. 2009b)]{Lanzaetal09b}, CoRoT-6 \cite[(Lanza et al. 2010a)]{Lanzaetal10a}, and CoRoT-7 \cite[(Lanza et al. 2010b)]{Lanzaetal10b}. From the migration of their active longitudes, a lower limit for the amplitude of their latitudinal differential rotation has been obtained. The results are listed in Table~\ref{lanza_table1}, together with those derived from MOST photometry and those by \cite[Mosser et al. (2009)]{Mosseretal09} for two of the CoRoT asteroseismic targets. In Table~\ref{lanza_table1}, the columns from left to right list the name of the star, its effective temperature $T_{\rm eff}$, its mean rotation period $P_{\rm rot}$, the relative amplitude of the differential rotation $\Delta \Omega / \Omega$, and the references. 

A comparison with the differential rotation amplitudes as derived from Doppler Imaging or the Fourier transform of the  spectral line profiles (\cite[Reiners 2006]{Reiners06}) shows that the values derived from the migration of the active longitudes are generally  smaller than those expected in  stars with the same effective temperature and rotation rates, by a typical factor of $2-3$. This suggests that their active regions are mostly localized at low latitudes, as in the case of the Sun. 
\begin{table}
  \begin{center}
  \caption{Stellar differential rotation from spot modelling of space-borne photometry.}
  \label{lanza_table1}
 {\scriptsize
  \begin{tabular}{|c|c|c|c|c|}
\hline 
Star & $T_{\rm eff}$ & $P_{\rm rot}$ & $\Delta \Omega / \Omega $ & References \\
  & (K) & (days) & & \\
\hline
$\epsilon $~Eridani & 4830 & 11.45 & $0.11 \pm 0.03^{*}$ & \cite[Croll et al. (2006)]{Crolletal06} \\
CoRoT-7 & 5275 & 23.64 & $0.058 \pm 0.017$ & \cite[Lanza et al. (2010b)]{Lanzaetal10b} \\
k$^{1}$~Ceti & 5560 & 8.77 & $0.09 \pm 0.006^{*}$ & \cite[Walker et al. (2007)]{Walkeretal07} \\
CoRoT-2 & 5625 & 4.52 & $\sim 0.009-0.08$ & \cite[Lanza et al. (2009a)]{Lanzaetal09a}; \\ 
& & & &  \cite[Fr{\"o}hlich et al.(2009)]{Frohlichetal09} \\
HD~175726 & 6030 & 3.95 & $\approx 0.40^{*}$ & \cite[Mosser et al. (2009)]{Mosseretal09} \\
CoRoT-6 & 6090 & 6.35 & $0.12 \pm 0.02$ & \cite[Lanza et al. (2010a)]{Lanzaetal10a} \\
CoRoT-4 & 6190 & 9.20 & $0.057 \pm 0.015$ & \cite[Lanza et al. (2009b)]{Lanzaetal09b} \\
HD~181906 & 6360 & 2.71 & $\approx 0.25^{*}$ & \cite[Mosser et al. (2009)]{Mosseretal09} \\ 
\hline
 \end{tabular}
  }
 \end{center}
\vspace{1mm}
 \scriptsize{
 {\it Note:}\\
  $^{*}$Value of the $K$ coefficient estimated for a solar-like differential rotation law $P(\phi) = P_{\rm eq}/(1-K \sin^{2} \phi)$, where $\phi $ is the latitude, $P(\phi)$ the rotation period at latitude $\phi$, and $P_{\rm eq}$ the rotation period at the equator. \\
  }
\end{table}
\section{The possible case for a magnetic star-planet interaction}
The planets of CoRoT-2, CoRoT-4, and CoRoT-6 are hot Jupiters, i.e., giant planets orbiting within 0.15 AU from their host stars. They interact  tidally and possibly magnetically with their stars, which may lead to observable effects on stellar activity  (see, e.g.,  \cite[Cuntz et al. 2000; Lanza 2008; 2009]{Cuntzetal00,Lanza08,Lanza09}). Current evidence of  star-planet magnetic interaction (hereafter SPMI) is  limited to the modulation of the chromospheric flux with the orbital phase of the planet in a few stars and  in some seasons \cite[(Shkolnik et al. 2005, 2008)]{Shkolniketal05,Shkolniketal08}. Evidence of a coronal flux enhacement is much more controversial (cf. \cite[Kashyap et al. 2008]{Kashyapetal08} and \cite[Poppenhaeger et al. 2010]{Poppenhaegeretal10}), although some possible cases have been presented \cite[(Saar et al. 2008; Pillitteri et al. 2010)]{Saaretal08,Pillitterietal10}.  SPMI features in the photosphere have been proposed for $\tau$ Bootis \cite[(Walker et al. 2008)]{Walkeretal08}, CoRoT-2 \cite[(Pagano et al. 2009)]{Paganoetal09}, and, possibly, HD~192263 \cite[(Santos et al. 2003)]{Santosetal03}. The mean rotation of $\tau$ Boo is synchronized  with the orbital motion of its giant planet, so a modulation of its optical flux with the orbital period of the planet cannot be unambiguously attributed to SPMI.
Nevertheless, \cite[Walker et al. (2008)]{Walkeretal08} found an active region on the star which lasted for at least $\sim 500$ stellar rotations, i.e., 5 years, always leading the subplanetary meridian by $\sim 70^{\circ}$. The persistence of such a feature  strongly suggests a connection with the planet. 

\cite[Lanza et al. (2009b)]{Lanzaetal09b} suggest a similar phenomenon in the other synchronous system CoRoT-4, finding an  active region located at the subplanetary longitude that has persisted for $\sim 70$ days. Even more intriguing is the case of CoRoT-6, a non-synchronous system with a  planetary orbital period of 8.886 days and a mean stellar rotation period of 6.35 days. Assuming a longitude reference frame rotating with the mean stellar rotation period, the  maximum of the spot filling factor in several  active regions occurs when they cross a meridian at $-200^{\circ}$  from the subplanetary meridian. The probability of a chance occurrence is only $\sim 0.8$ percent (\cite[Lanza et al. 2010a]{Lanzaetal10a}). 

It is difficult to find a mechanism for the allegedly supposed influence of the planet on the formation of stellar active regions. \cite[Lanza (2008)]{Lanza08}  conjectured that the reconnection of the stellar coronal field with the magnetic field of the planet may induce a longitudal dependence of the hydromagnetic dynamo action in the star, 
 provided that some of the spot magnetic flux tubes come from the subphotospheric layers, as suggested by \cite[Brandenburg (2005)]{Brandenburg05}. Nevertheless, further observations of the photospheric SPMI are needed firstly to confirm the reality of the phenomenon and secondly to derive its dependence on stellar and planetary parameters.


\begin{thebibliography}{}


\bibitem[Aigrain et 
al.(2008)]{Aigrainetal08} Aigrain, S., et al.\ 2008, \textit{A\&A}, 488, L43 

\bibitem[Alonso et 
al.(2008)]{Alonsoetal08} Alonso, R., et al.\ 2008, \textit{A\&A}, 482, L21 



\bibitem[Auvergne et 
al.(2009)]{Auvergneetal09} Auvergne, M., et al.\ 2009, \textit{A\&A}, 506, 411 

\bibitem[Berdyugina(2005)]{Berdyugina05} Berdyugina, S.~V.\ 2005, 
\textit{Living Reviews in Solar Physics}, 2, 8 

\bibitem[Berdyugina \& Tuominen(1998)]{BerdyuginaTuominen98} 
Berdyugina, S.~V., \& Tuominen, I.\ 1998, \textit{A\&A}, 336, L25 


\bibitem[Bouchy et 
al.(2008)]{Bouchyetal08} Bouchy, F., et al.\ 2008, \textit{A\&A}, 482, L25 

\bibitem[Brown et 
al.(2010)]{Brownetal10} Brown, A., Korhonen, H., Berdyugina, S, et al., these proceedings 


\bibitem[Brandenburg(2005)]{Brandenburg05} Brandenburg, A.\ 2005, 
\textit{ApJ}, 625, 539 

\bibitem[Ciardi et al. (2010)]{Ciardietal10} Ciardi, D.~R., von 
Braun, K., Bryden, G., et al.\ 2010, \textit{ApJ}, submitted, arXiv:1009.1840 



\bibitem[Collier Cameron(1997)]{CollierCameron97} Collier Cameron, A.\ 
1997, \textit{MNRAS}, 287, 556 

\bibitem[Croll et al.(2006)]{Crolletal06} Croll, B., et al.\ 2006, 
\textit{A\&A}, 648, 607 

\bibitem[Cuntz et al.(2000)]{Cuntzetal05} Cuntz, M., Saar, S.~H., 
\& Musielak, Z.~E.\ 2000, \textit{ApJ}, 533, L151 


\bibitem[Dupree (2010)]{Dupree10}
{Dupree, A.} 2010, these proceedings

\bibitem[Fr{\"o}hlich et 
al.(2009)]{Frohlichetal09} Fr{\"o}hlich, H.-E., K{\"u}ker, M., Hatzes, A.~P., \& Strassmeier, K.~G.\ 2009, \textit{A\&A}, 506, 263 



\bibitem[Fr{\"o}hlich 
\& Lean(2004)]{FrohlichLean04} Fr{\"o}hlich, C., \& Lean, J.\ 2004, \textit{A\&AR}, 12, 273 



\bibitem[Haas et al.(2010)]{Haasetal10} Haas, M.~R., et al.\ 2010, 
\textit{ApJ}, 713, L115 

\bibitem[Huber et 
al.(2010)]{Huberetal10} Huber, K.~F., Czesla, S., Wolter, U., \& Schmitt, J.~H.~M.~M.\ 2010, \textit{A\&A}, 514, A39 


\bibitem[Kashyap et al.(2008)]{Kashyapetal08} Kashyap, V.~L., Drake, 
J.~J., \& Saar, S.~H.\ 2008, \textit{ApJ}, 687, 1339 



\bibitem[Koll{\'a}th 
\& Ol{\'a}h(2009)]{KollathOlah09} Koll{\'a}th, Z., \& Ol{\'a}h, K.\ 2009, \textit{A\&A}, 501, 695 


\bibitem[Krivova 
\& Solanki(2002)]{krivovaSolanki02} Krivova, N.~A., \& Solanki, S.~K.\ 2002, \textit{A\&A}, 394, 701 


\bibitem[Lanza(2008)]{Lanza08} Lanza, A.~F.\ 2008, \textit{A\&A}, 487, 1163 


\bibitem[Lanza(2009)]{Lanza09} Lanza, A.~F.\ 2009, \textit{A\&A}, 505, 339 


\bibitem[Lanza et 
al.(1998)]{Lanzaetal98} Lanza, A.~F., Catalano, S., Cutispoto, G., Pagano, I., \& Rodono, M.\ 1998, \textit{A\&A}, 332, 541 

\bibitem[Lanza et 
al.(2007)]{Lanzaetal07} Lanza, A.~F., Bonomo, A.~S., \& Rodon{\`o}, M.\ 2007, \textit{A\&A}, 464, 741 

\bibitem[Lanza et 
al.(2009a)]{Lanzaetal09a} Lanza, A.~F., et al.\ 2009a, \textit{A\&A}, 493, 193 




\bibitem[Lanza et 
al.(2009b)]{Lanzaetal09b} Lanza, A.~F., et al.\ 2009b, \textit{A\&A}, 506, 255 


\bibitem[Lanza et al.(2010a)]{Lanzaetal10a} Lanza, A.~F., et al.\ 
2010a, \textit{A\&A}, in press, arXiv:1007.3647 


\bibitem[Lanza et al.(2010b)]{Lanzaetal10b} Lanza, A.~F., et al.\ 
2010b, \textit{A\&A}, in press, arXiv:1005.3602 



\bibitem[Lou(2000)]{Lou00} Lou, Y.-Q.\ 2000, \textit{ApJ}, 540, 1102 


\bibitem[Messina 
\& Guinan(2003)]{MessinaGuinan03} Messina, S., \& Guinan, E.~F.\ 2003, \textit{A\&A}, 409, 1017 


\bibitem[Mosser et 
al.(2009)]{Mosseretal09} Mosser, B., Baudin, F., Lanza, A.~F., et al.\ 2009, \textit{A\&A}, 506, 245 



\bibitem[Ol{\'a}h (2010)]{Olah10} Ol{\'a}h, K. 2010, these proceedings


\bibitem[Oliver et al.(1998)]{Oliveretal98} Oliver, R., Ballester, 
J.~L., \& Baudin, F.\ 1998, \textit{Nature}, 394, 552 

\bibitem[Pagano et al. (2009)]{Paganoetal09} Pagano, I., Lanza, A.~F., Leto, G., et al.\ 2009, \textit{Earth, Moon,  and Planets}, 105, 373 



\bibitem[Pillitteri et al.(2010)]{Pillitterietal10} Pillitteri, I., 
Wolk, S.~J., Cohen, O., et al. \ 2010, \textit{ApJ}, in press, arXiv:1008.3566 


\bibitem[Poppenhaeger et 
al.(2010)]{Poppenhaegeretyal10} Poppenhaeger, K., Robrade, J., \& Schmitt, J.~H.~M.~M.\ 2010, \textit{A\&A}, 515, A98 


\bibitem[Reiners(2006)]{Reiners06} Reiners, A.\ 2006, \textit{A\&A}, 446, 267 

\bibitem[Saar et al.(2008)]{Saaretal08} Saar, S.~H., Cuntz, M., 
Kashyap, V.~L., \& Hall, J.~C.\ 2008, IAU Symposium, 249, 79 

\bibitem[Santos et al. (2003)]{Santosetal03} Santos, N.~C., et al.\ 2003, \textit{A\&A}, 406, 373 



\bibitem[Savanov(2010)]{Savanov10} Savanov, I.~S.\ 2010, 
\textit{Astron. Rep.}, 54, 437 




\bibitem[Sch{\"u}ssler 
\& Rempel(2005)]{SchusslerRempel05} Sch{\"u}ssler, M., \& Rempel, M.\ 2005, \textit{A\&A}, 441, 337 


\bibitem[Shkolnik et al.(2005)]{Shkolniketal05} Shkolnik, E., et al.\ 2005, \textit{ApJ}, 622, 1075 



\bibitem[Shkolnik et al.(2008)]{Shkolniketal08} Shkolnik, E., 
Bohlender, D.~A., Walker, G.~A.~H., 
\& Collier Cameron, A.\ 2008, \textit{ApJ}, 676, 628 

\bibitem[Silva-Valio et al.(2010)]{SilvaValio10} 
Silva-Valio, A. 2010, these proceedings


\bibitem[Silva-Valio et al.(2010)]{SilvaValioetal10} 
Silva-Valio, A., Lanza, A.~F., Alonso, R., \& Barge, P.\ 2010, \textit{A\&A}, 510, A25 



\bibitem[Strassmeier (2009)]{Strassmeier09}
{Strassmeier, K. G.,} 2009, \textit{A\&AR}, 17, 251

\bibitem[Strassmeier (2010)]{Strassmeier10}
{Strassmeier, K. G.,} 2010, these proceedings

\bibitem[Walker et al.(2003)]{Walkeretal03} Walker, G., et al.\ 2003, \textit{PASP}, 115, 1023 

\bibitem[Walker et al.(2007)]{Walkeretal07} Walker, G.~A.~H., et 
al.\ 2007, \textit{ApJ}, 659, 1611 

\bibitem[Walker et 
al.(2008)]{Walkeretal08} Walker, G.~A.~H., et al.\ 2008, \textit{A\&A}, 482, 691 

\bibitem[Winn et al.(2005)]{Winnetal05} Winn, J.~N., et al.\ 2005, 
\textit{ApJ}, 631, 1215 


\bibitem[Wolter et al.(2009)]{Wolteretal09} Wolter, U., Schmitt, J.~H.~M.~M., Huber, K.~F., et al.\ 2009, \textit{A\&A}, 504, 561 

\bibitem[Zappal\`a 
\& Zuccarello (1991)]{ZappalaZuccarello91} Zappal\`a, R.~A., \& Zuccarello, F.\ 1991, \textit{A\&A}, 242, 480 

\bibitem[Zaqarashvili et al.(2010)]{Zaqarashvilietal10} Zaqarashvili, 
T.~V., Carbonell, M., Oliver, R., \& Ballester, J.~L.\ 2010, \textit{ApJ}, 709, 749 





\end{thebibliography}
\end{document}